\documentclass{osa-article}

\journal{osajournal}

\articletype{Research Article}

\begin{document}

\title{Stress measurements in silicon photonics by integrated Raman spectroscopy}

\author{Guillaume Marcaud,\authormark{1,*} Mathias Berciano,\authormark{1} Christian Lafforgue,\authormark{1} Carlos Alonso-Ramos,\authormark{1} Xavier Le Roux,\authormark{1} Thomas Maroutian,\authormark{1} Guillaume Agnus,\authormark{1} Pascal Aubert,\authormark{1} Ludovic Largeau,\authormark{1} Eric Cassan,\authormark{1} Sylvia Matzen,\authormark{1} Delphine Marris-Morini,\authormark{1} Philippe Lecoeur,\authormark{1} and Laurent Vivien\authormark{1,$\dagger$}}

\address{\authormark{1}Universit\'e Paris-Saclay, CNRS, Centre de Nanosciences et Nanotechnologies (C2N),UMR 9001, Palaiseau, 91120, France}

\email{\authormark{*}guillaume.marcaud@yale.edu\\\authormark{$\dagger$}laurent.vivien@c2n.upsaclay.fr} 



\begin{abstract*}
Complex 3D integration of photonic and electronic integrated circuits is of particular interest to carry the photonics roadmap and to address challenges but involves mechanical stress, often detrimental for the behavior of optical components. Existing experiments failed to carefully analyze the stress in such integrated optical devices due to the requirement in terms of feature sizes, few hundreds of nanometers, and 3D-stacked integration. We present for the first time the characterization of the stress tensor of a silicon waveguide using Integrated Raman Spectroscopy (IRS). This experimental technique is directly sensitive to the effective stress, which involves changes in optical properties of the guided mode, at the working wavelength and polarization state of the photonic component. The experimental stress tensor is in good agreement with simulations.
\end{abstract*}

\section{Introduction}
Silicon photonics has been identified as the solution to improve the telecommunication networks, data centers facilities and computer systems while preserving low power consumption and tremendous data rates. This field of research has exhibited an impressive development the last decade, driven by the maturity of the silicon platform to integrate closely and at the wafer-scale, photonic and electronic circuits. Further bandwidth improvement towards Tb/s and the increase of the number of on-chip functionalities, requires complex integration schemes including 3D partitioning and interconnects \cite{coudrain_towards_2012}. One promising solution for that purpose is the use of Through-Silicon Via (TSV), consisting in a copper vertical interconnections crossing the silicon substrate.  However, the implementation of 3D interconnects involved mechanical stress \cite{vianne_through-silicon_2015} which can be critical for the performances and reliability \cite{fage-pedersen_linear_2006}.

In this context, probing the stress in complex sub-micrometer devices becomes an important challenge, vital for design optimization and development of 3D stacked structures. Among the tools available, micro-Raman spectroscopy, widely used to characterize stress in silicon, presents several advantages. First, the technique is non-destructive and very fast. Second, thanks to the recent technological progress of lasers and detectors, Raman systems are very accurate spectrally and spatially. Depending of the wavelength source and instrument calibration, the accuracy of stress measurements can reach few tens of MPa. Even if the relationships between experimental measurements and stress fields are not straightforward, the theoretical framework is known and has been experimentally validated for simple systems \cite{de_wolf_relation_2015,anastassakis_effect_1970,wolf_stress_1996,ganesan_lattice_1970}.

However, conventional Backscattering Raman Spectroscopy (BRS) is not convenient to address the evolution of  photonic and electronic circuits. Indeed, Raman systems probe samples from the surface, which is not possible when silicon is buried or encapsulated by a multi-stack cladding including metal plugs, oxides and nitrides. Furthermore, this technique suffers from its incapacity to resolve the tensor nature of the stress due to its intrinsic configuration, which allow to probe typically only one phonon. Silicon possesses three Raman-active optical phonons and all the information regarding the stress is contained in their frequency and intensity. Therefore, if only one phonon is revealed, only a partial information can be obtained. The way to observe the three phonons is to probe the silicon with an incident and scattered optical electric field that present a component in each of the three crystallographic axes. This condition is not permitted with the z=[001] incident direction of the laser in BRS experiments \cite{anastassakis_effect_1970} but has been achieved in several other studies either by tilting away from the [001] axis the incident light beam \cite{loechelt_measuring_1995}, or more recently by using BRS equipped with high numerical aperture lenses \cite{poborchii_observation_2010}. Although those methods have demonstrated promising results, they are still not adapted for complex 3D integrated photonic circuits and give only a local information about the stress in the component. Another approach consists of performing BRS on the $\langle 110 \rangle$ cross sections of a cleaved samples \cite{poborchii_study_2007}, which is destructive and not achievable for very small samples or nanometer-sized components.

\begin{figure}
\centering
\includegraphics[scale=0.3]{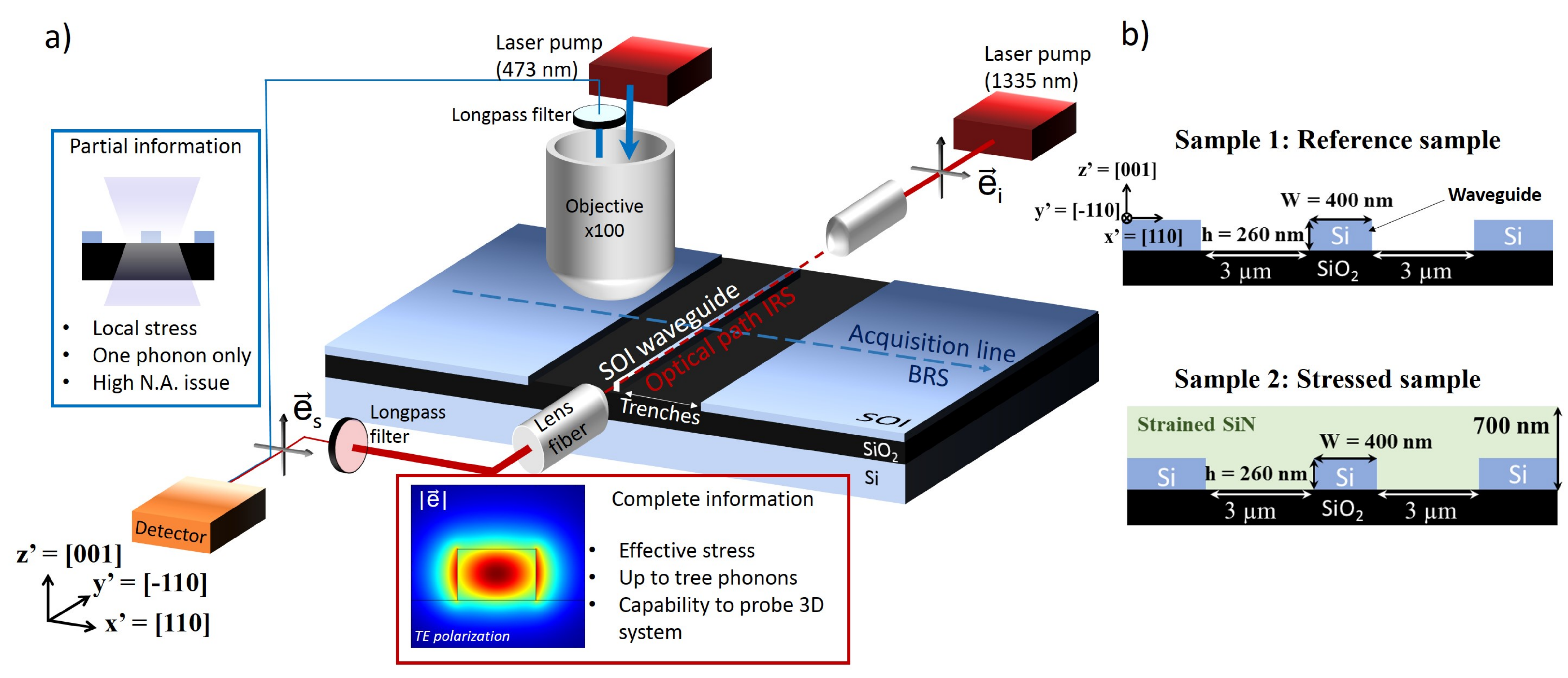}
\caption{\label{fig:RAMAN} Raman spectroscopy set-up combining IRS and BRS configurations. BRS experiments are performed with a $\lambda$=473 nm laser pump incident along the z'=[001] direction of silicon and only one phonon is typically probed. This technique is coupled to a mobile stage allowing the mapping of the device. IRS experiments exploit the confinement of a 1335 nm laser pump into a y'=[1-10] oriented monomode waveguide. The incident and scattered polarization allow to probe up to three phonons.}
\end{figure}

To overcome the limitations of conventional technique, we develop a new Integrated Raman Spectroscopy (IRS). In this configuration, the laser pump is directly injected into a silicon waveguide and the Raman signals is collected at the output. Whereas Raman emission in a waveguide has been already observed \cite{claps_observation_2002} and used for different applications such as detection of biological molecules \cite{holmstrom_trace_2016,zhao_stimulated_2018,dhakal_nanophotonic_2016}, Raman signal enhancement \cite{wang_surface_2016,wong_nanoscale_2018} or thin film characterization \cite{duverger_waveguide_1999}, the non-degenerates optical phonons of silicon have never been exploited to study the stress tensor. Benefiting from the strong light confinement in integrated optical components, IRS is more efficient to characterize the stress in nanometer-sized structures, whatever the cladding and structure geometry. Moreover, IRS allows to directly probe the stress-field convoluted with the optical mode at the working wavelength and polarization state of the components, which is perfectly adapted to study any stress-induced change of silicon properties.

In this study, the stress has been determined for two 400 nm wide strip silicon waveguides fabricated in a 260 nm thick Silicon-On-Insulator (SOI) substrate. As represented in Figure \ref{fig:RAMAN} b), the sample 1 is a Si-waveguide without top cladding, used as a reference while the sample 2 is covered by 700 nm of compressively strained-SiN. The deposition process ensures a biaxial stress of about -1.3 GPa in-plane in the SiN layer, measured by the curvature radius method on a ultra-flat silicon wafer. Both samples 1 and 2 are characterized by Raman spectroscopy with BRS and IRS configurations, whose a schematic view is given Figure \ref{fig:RAMAN} a). The results are presented as follow. First, the spectra acquired in IRS and BRS confguration are compared and discussed. Then, based on Finite Element Method (FEM) simulations, a theoretical framework is built, allowing to extract the effective stress tensor from the phonons frequencies measured in IRS. Finally, the possibility to study the evolution of each phonon as a function of the output light polarization is presented as a promising perspective of this work. In the next, the quotes denotes quantities given in the device axes system \{x'=[110], y'=[-110], z'=[001]\}, in opposition of the crystallographic axes system \{x=[100], y=[010], z=[001]\}.

\section{Experimental results}

\begin{figure}
\centering
\includegraphics[scale=0.45]{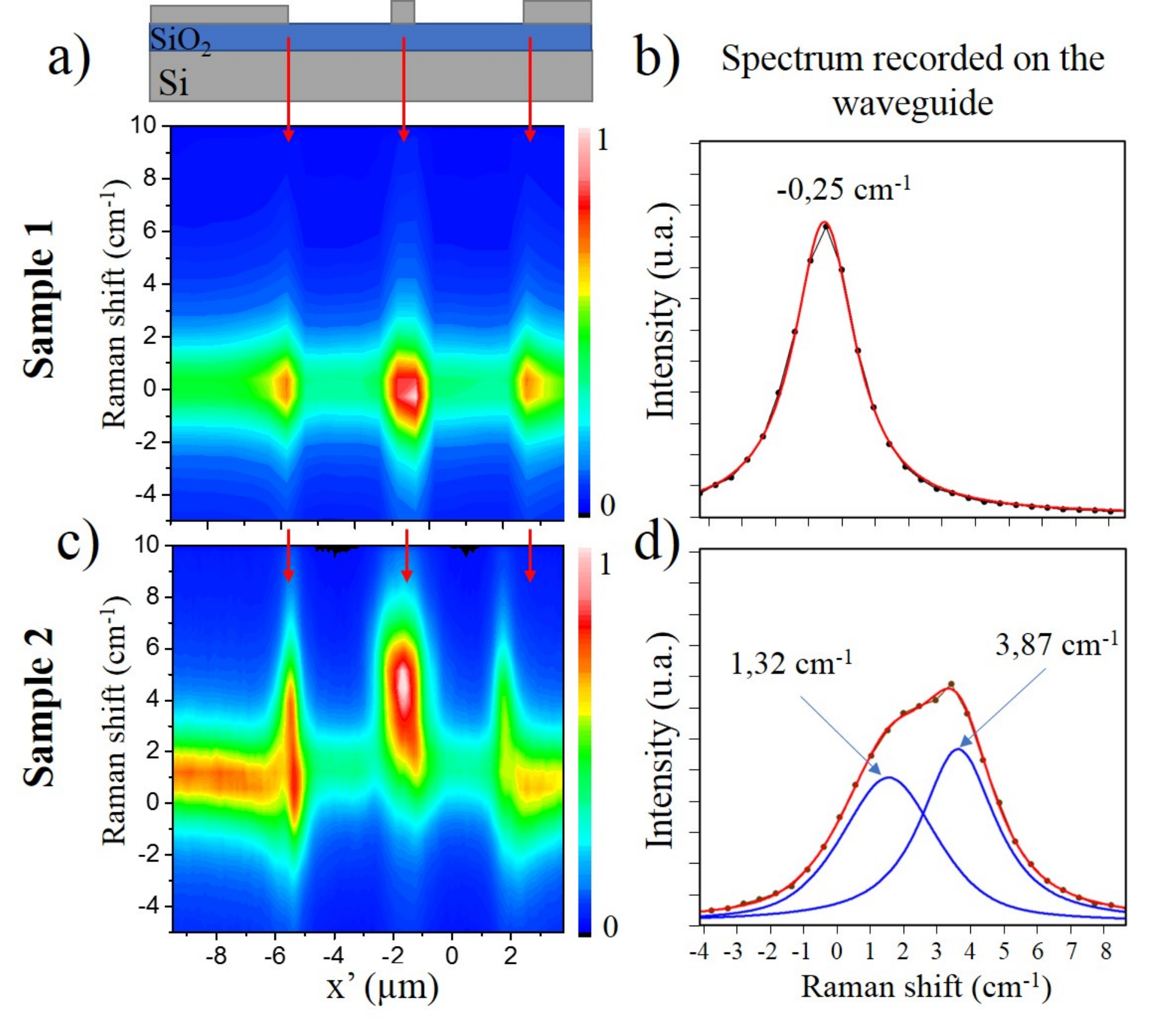}
\caption{\label{fig:ligneScan} Raman maps in BRS configuration for the reference a) and stressed sample c). The spectrum acquired on top of the waveguide is also displayed on the right. Only one Voigts function fits the reference spectrum in b), and two functions are needed for the stressed waveguide in d). A schematic view of the device cross-section is also included.}
\end{figure}

The Raman 1D mapping Figure \ref{fig:ligneScan} a) and c) for the sample 1 and 2 respectively, are performed with BRS at a laser wavelength of $\lambda$=473 nm and along the x' axis, perpendicular to the waveguides. A schematic cross-section of the devices is also presented in order to associate the x' coordinates of the maps to the different parts of the waveguide. We observed for both structures a broadening of the Raman peak as well as an increase of the signal intensity at the edges of the structure, indicated by red arrows: at both sides of the trenches and on the waveguide. Whereas the intensity increase is likely due to the polarization dependent signal enhancement from the edge discontinuity of the refractive index \cite{poborchii_observation_2010}, the broadening may be due to the raise of a second Raman mode and/or due to larger stress gradients. The spectra in Figure \ref{fig:ligneScan} b) and d) correspond to the ones recorded on the top of the waveguide. The spectrum obtained on the reference waveguide presents only one peak. Its position is determined by fitting the peak with a Voigt function centered at $\Delta \omega^{BRS}=-0.25$ $\text{cm}^{-1}$, where the Raman shift $\Delta \omega_i=\omega_i-\omega_0$ is calculated from our reference phonon frequency of silicon at $\omega_0=520.7$ $\text{cm}^{-1}$. The spectrum recorded on the stressed waveguide presents a clear doubling of the Raman peak. Two functions are needed to correctly fit the spectrum at $\Delta \omega^{BRS}=+1.32$ $\text{cm}^{-1}$ and $\Delta \omega^{BRS}=+3.87$ $\text{cm}^{-1}$.

In order to understand the spectral signature of both samples, we have to look at the selection rules of the Raman process involved \cite{wolf_stress_1996, de_wolf_relation_2015,ganesan_lattice_1970,anastassakis_effect_1970}. The latter depends on the incident ($\boldsymbol{e_i}$) and scattered ($\boldsymbol{e_s}$) polarization orientation with respect to the silicon crystallographic axes such as $I_j=I_0|\boldsymbol{e_i} \cdot \boldsymbol{R_j} \cdot \boldsymbol{e_s}|^2$, where $I_j$ and $\boldsymbol{R_j}$ are the intensity and the Raman tensor of the phonon $j$. The typical results of the selection rule is presented Table \ref{tab:selection}. In an ideal BRS experiment, where the incident and scattered light is only along the [001] direction, the selection rules predict that only the third phonon of silicon can be observed. However, we clearly observe two components in the Raman spectrum of the sample 2 (strained sample). This doubling of the Raman peak is due to a misalignment of the incident and/or scattered light with the [001] axis that allow a second phonon mode to be observed and a strain state in the silicon lattice that lifts the degeneracy of the phonons frequency. On the one hand, the incident and/or scattered light in our BRS setup is likely misaligned because of the the use of a high Numerical Aperture objective (objective x100, N.A.=0.9). The wide emission and collection angle allowed by this objective, coupled to the strong refractive index and the nanometric dimension of the silicon waveguide, imply that the incident and scattered light include a non negligible component perpendicular to the [001] silicon axis. On the other hand, one can have a misalignement of the [001] crystallographic axis itself with respect to the sample normale. Even though this phenomenon is usually weak, it can be stronger in presence of high shear stress involving the [001] axis in nanometric structures \cite{wolf_stress_1996}.

\begin{table}[h]
\centering
    \caption{\label{tab:selection} Raman selection rules considered for BRS and IRS configurations for typical incident $\boldsymbol{e}_i$ and scattered $\boldsymbol{e}_s$ polarizations directions. The intensites are function of $d$, the Raman polarizability \cite{aggarwal_measurement_2011}.}
    \small
        \begin{tabular}{|c|c|c|c|c|c|}
        \hline
        Configuration & \hspace{2mm}  $\boldsymbol{e}_i$ \hspace{2mm} & \hspace{2mm}  $\boldsymbol{e}_s$ \hspace{2mm}  & \hspace{2mm}  $I_1$ \hspace{2mm}  & \hspace{2mm}  $I_2$ \hspace{2mm}  &  \hspace{2mm} $I_3$ \hspace{2mm} \\
        \hline
        \hline
        &  &   &   &   &\\
       \textbf{BRS} & x' & x' & 0 & 0 & $d^2$  \\
       \textbf{configuration} & x' & y' & 0 & 0 & 0 \\
       &   &    &   &   &\\
        \hline
        &   &    & & &\\
        \textbf{IRS} & x' & x' & 0 & 0 & $d^2$\\
       \textbf{configuration}  & x' & z' & $d^2$/2 & $d^2$/2 & 0 \\
        &  &   &   &   &\\
        \hline
        \end{tabular}
\end{table}

\begin{figure}
\centering
\includegraphics[scale=0.22]{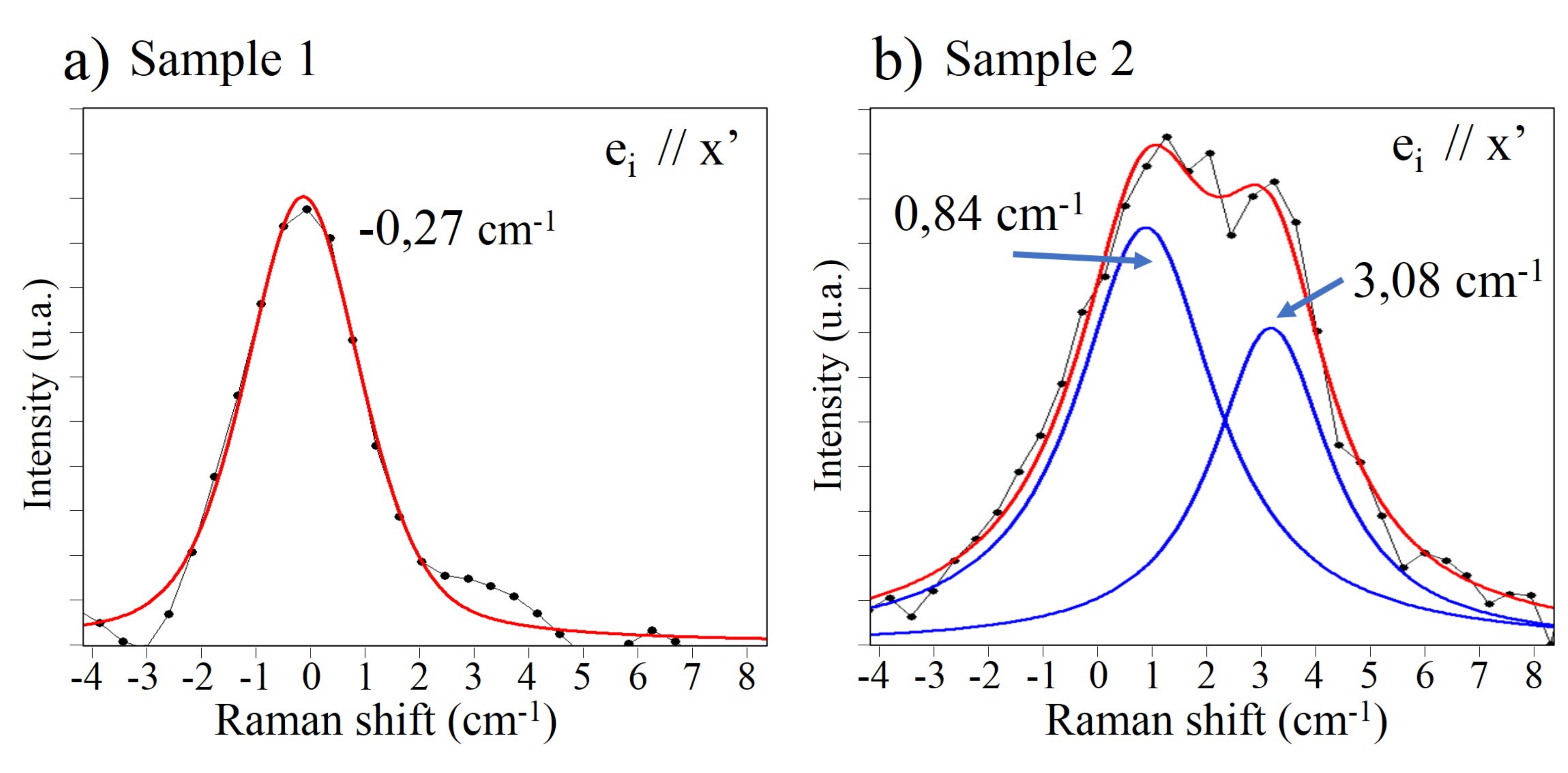}
\caption{\label{fig:IRS} Raman spectra recorded by IRS with a x'-polarised laser pump ($e_i$=x', quasi-TE optical mode)  at $\lambda$=1335 nm and free scattered polarization ($e_s$). The Raman peak is fitted with one or two Voigt functions for the reference a) and stressed sample b), respectively.}
\end{figure}

The same waveguides are then characterized with the IRS setup. A laser pump at a wavelength of $\lambda=$1335 nm in TE polarization ($e_i$ = x') is injected into the Si waveguide and no polarizer was used at the waveguide output to record the Raman signal. The resulting spectra are displayed Figure \ref{fig:IRS} a) and b) for the sample 1 and 2, respectively. Only one Raman peak is obtained for the sample 1 with a fitting peak located at $\Delta \omega^{IRS}=-0.27$ $\text{Rcm}^{-1}$, very similar to the frequency measured with BRS. However, the three phonons are allowed (Table \ref{tab:selection}) in this experiment. This results confirms the degenerate nature of the three phonons in this reference waveguide. Now that we know the frequency of the three phonons, one can resolve the system given by the secular equation and directly compute, without fitting parameter, up to three components of the stress tensor following the method described here \cite{de_wolf_relation_2015}. Considering a triaxial model where only the diagonal components of the stress tensor are non null, one can found an effective stress $\sigma_{11}^{\prime eff}$ = $\sigma_{22}^{\prime eff}$ = $\sigma_{33}^{\prime eff}$=47 MPa. This stress is considered effective as it is the result of the convolution of the optical mode of the pump used in IRS and the stress field in the waveguide. These values are in agreement with the literature for a fabrication-induced stress \cite{camassel_strain_2000,wolf_stress_1999} in our reference sample.

\begin{figure}
\centering
\includegraphics[scale=0.34]{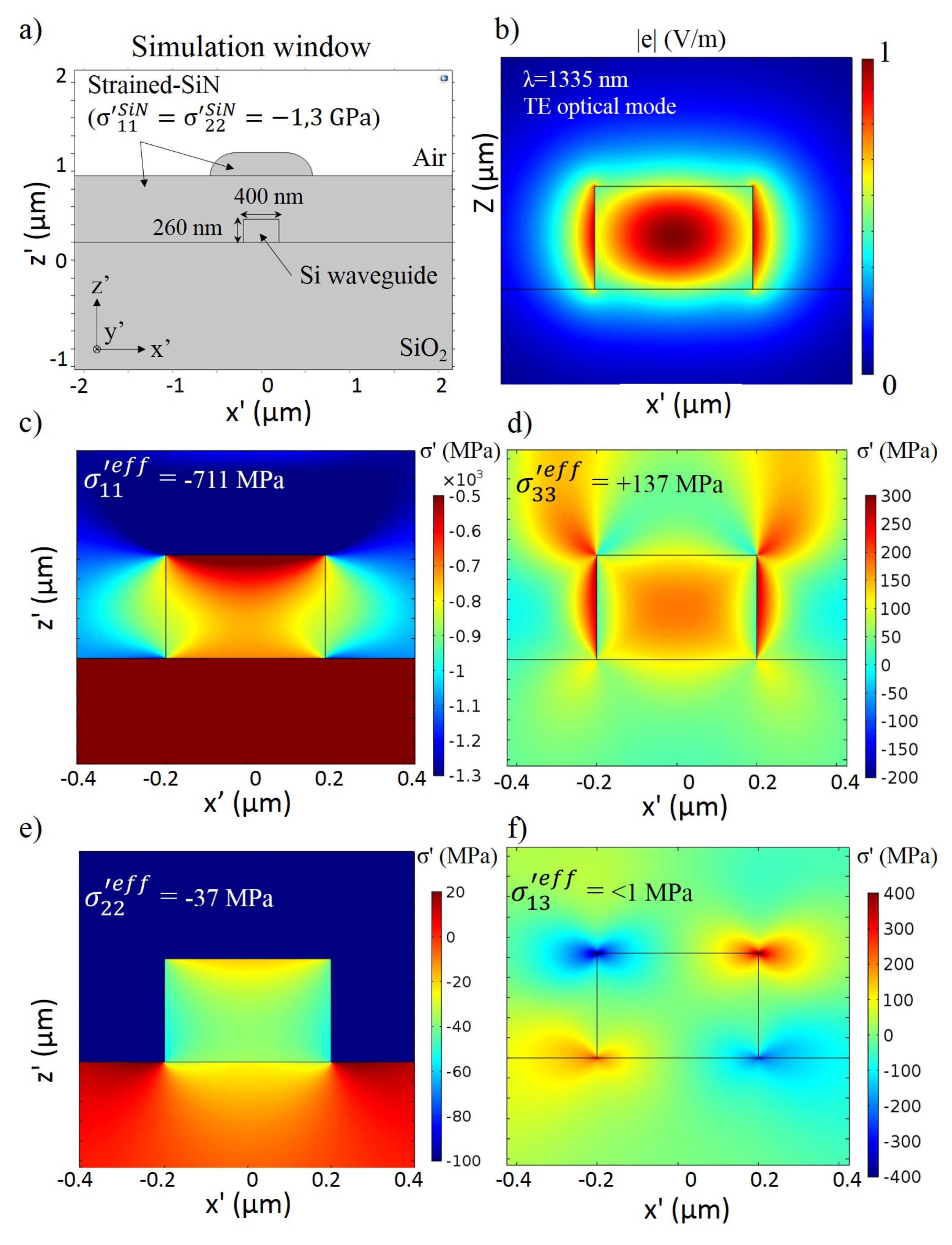}
\caption{\label{fig:Simu} Stress and optical mode confinement simulation. a) Dimensions of the waveguide and simulation window. b) Optical quasi-TE mode confined in the waveguide at $\lambda=1335 nm$. c) to f) Stress field distribution of the four non-zero components of the stress tensor. The effective stress values, considering the overlap with the electric field distribution of the optical mode, are indicated on the top of each respective window.}
\end{figure}

The spectrum acquired throughout the strained-silicon waveguide presents a wide and asymmetric Raman peak. Keeping about the same Full Width at Half Maximum (FWHM) of the fitting peaks used in BRS, the IRS spectrum can be efficiently fitted with two Voigt functions centered at $\Delta \omega^{IRS}=+0.84$ $\text{cm}^{-1}$ and $\Delta \omega^{IRS}=+3.08$ $\text{cm}^{-1}$. The frequencies are also very similar to the ones measured in BRS, confirming the robustness of our IRS setup. Once again, to determine up to three components of the effective stress tensor in the waveguide of the sample 2, assumptions have to be made. However, as we discussed previously, the raise of a second phonon in BRS may be due to a strong shear stress in the waveguide, in disagreement with the triaxial stress model used for the reference sample. In order to shine light on the expected stress in this system, we have performed Finite Element Method simulations (FEM). The simulated stress fields are presented Figures \ref{fig:Simu} c)-f) for the non-null components of the stress tensor. As expected, the stress in the silicon waveguide is mainly compressive along the x' direction $\sigma_{11}^\prime$, perpendicular to the waveguide. At a lower level, the stress $\sigma_{22}^\prime$ is also compressive and $\sigma_{33}^\prime$ tensile. The shear stress $\sigma_{13}^\prime$ is weak except in the corners of the waveguide where it reaches few hundreds of MPa. As the FEM simulations are performed in 2D, the others shear stress $\sigma_{23}^\prime$ and $\sigma_{12}^\prime$ cannot be determined but they are also expected to be negligible due to the infinite length of the waveguide in the y' direction. The stress probed by IRS technique is sensitive to the electric-field distribution of the optical mode in the structure. To take into account this contribution, we have also simulated the quasi-TE mode of the laser pump at $\lambda$=1335 nm, Figure \ref{fig:Simu} b), propagating into the silicon waveguide. Combining the stress ($\sigma^{\prime}$) and electric fields ($|\vec{e}|$) distribution of the optical mode, we are able to extract the effective stress in the waveguide $\sigma^{\prime eff}_{ij}=\sigma_{ij}^\prime*|\vec{e}|/\int_{wg} |\vec{e}| $, figure of merit of the component, as it is directly related to the evolution of its optical property. For the diagonal components, the effective values reported are $\sigma_{11}^{\prime eff}$=-711 MPa,  $\sigma_{22}^{\prime eff}$=-37 MPa and  $\sigma_{33}^{\prime eff}$=+137 MPa. The shear stress $\sigma_{13}$ presents positive or negative values and is mainly localized at the waveguide corners. Its overlap with the optical mode is very weak $\sigma_{13}^{\prime eff}$=$<$1 MPa.

\begin{figure}
\centering
\includegraphics[scale=0.35]{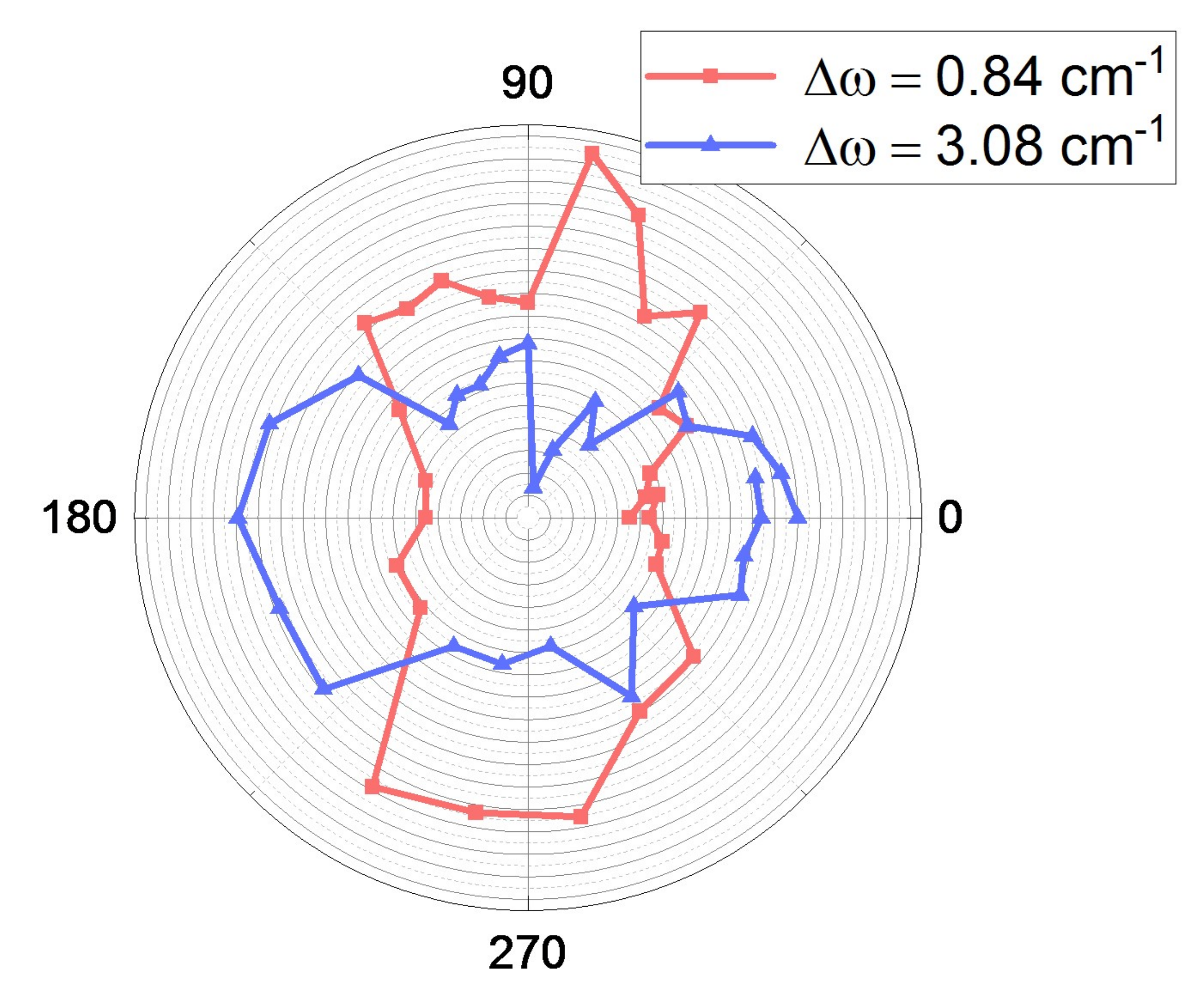}
\caption{\label{fig:Polar} Experimental areas of the two Voigt components used to fit the IRS spectra in function of the scattered polarization $e_s$. The incident laser pump polarization is kept along the $e_i$=x' direction. When $e_s=0^\circ$, the scattered polarization is along the x' direction.}
\end{figure}

From the simulations, it is clear that the effective shear stress is negligible with respect to the diagonal terms. In this picture, the less restrictive set of assumptions fall into the triaxial model. However, in order to compute the three components of the effective stress tensor, one has to attribute the two experimentally measured frequencies to the three phonons, and therefore consider a two-fold degeneracy. The calculation of the phonons frequencies from the simulations results gives an insight into this attribution. Indeed, the three diagonal terms of the simulated effective stress tensor lead to $\Delta \omega^{}_1 \approx \Delta \omega^{}_3= 1.8$ $\text{cm}^{-1}$ and $\Delta \omega^{}_2=0.4$ $\text{cm}^{-1}$. Even if the computed Raman shifts are far from the experimental results, we note that the two degenerated phonons in IRS are most likely the phonon 1 and 3. Another way to study the validity of this attribution is to look at the intensity of the two fitting components as a function of the scattered polarization ($e_s$), see Figure \ref{fig:Polar}. 
Following the same kind of degeneracy, we attribute the experimental Raman shift to the following phonons mode $\Delta \omega^{IRS}_1 = \Delta \omega^{IRS}_3= 3.08$ $\text{cm}^{-1}$ and $\Delta \omega^{IRS}_2=0.84$ $\text{cm}^{-1}$, which lead to  $\sigma_{11}^{\prime eff}$=-1202 MPa,  $\sigma_{22}^{\prime eff}$=-248 MPa and  $\sigma_{33}^{\prime eff}$=+227 MPa. Note that any other attributions lead to incoherent results. The experimental results follow the same trend as the simulations but differ in amplitude as presented table \ref{tab:ResultatContrainte}.

\begin{table}{}
\centering
\caption{\label{tab:ResultatContrainte} Experimental and simulated values of the effective stress in the silicon waveguide of sample 2. The experimental stress components have been calculated from the three Raman mode frequencies measured in IRS configuration and the simulated values are given for two different level of biaxial stress in the SiN layer.}
\begin{tabular}{|c|c|c|c|}
\hline
Componnents & Experiments & Simulation & Simulation \\
& (IRS) & \footnotesize{$\sigma^{\prime SiN}$=-1.2 GPa} & \footnotesize{$\sigma^{\prime SiN}$=-2.2 GPa} \\
\hline 
\hline 
&  &  &\\
$ \sigma_{11}^{\prime eff}$ & -1202 $\pm 6 $ MPa & -711 MPa & -1203 MPa \\
&  &  &\\
$ \sigma_{22}^{\prime eff}$ & -248 $\pm 23 $ MPa & -37 MPa & -62 MPa\\
&  &  &\\
$\sigma_{33}^{\prime eff}$ & +227 $\pm 4 $ MPa & +137 MPa & +232 MPa \\
&  &  &\\
$\sigma_{13}^{\prime eff}$ & x & $<$1 MPa & $<$1 MPa \\
&  &  &\\
\hline
\end{tabular}
\end{table}

The difference between experiments and simulations can be understood by considering a higher level of stress in the SiN layer. Indeed, simulations performed with a biaxial stress of -2.2 GPa instead of -1.3 GPa in the SiN layer shows results closer to experiments with $\sigma_{11}^{\prime eff}$=-1203 MPa,  $\sigma_{22}^{\prime eff}$=-62 MPa and  $\sigma_{33}^{\prime eff}$=+232 MPa (see table \ref{tab:ResultatContrainte}). Nevertheless, the stress in the SiN layer, determined from the curvature method after a deposition on an ultra-flat 3 inches silicon wafer, ensures a biaxial in-plane stress of $-1.3 \pm 0.1$  GPa. First, this value is determined within the assumptions of the curvature method and therefore can differs in some extents from the actual stress in the SiN layer. Secondly, the 1x2 $cm^2$ SOI patterned sample, covered with hundreds of photonic components, should behave differently and influence the stress resulting in the SiN layer during deposition. Indeed, the geometry of the sample, the fully etched waveguide, and the interface of the top silicon (260 nm) with the amorphous silica, may play a role in the internal stress of SiN and its transfer into the silicon.\\

Considering a higher stress in the SiN about 2.2 GPa, the experimental results are very close to the simulations and give an interesting insight into the effective stress felt by the optical mode in the waveguide.

Another interesting aspect of IRS is the use of the scattered polarization $e_s$ as another degree of freedom. For instance it can be used to check the validity of the frequencies attribution to the three phonons. The following experiment does not aim to prove the degeneracy of the phonons 1 and 3 but presents an interesting opening into the potential of IRS to study separately the three phonons of silicon. Whereas the incident laser polarization is still kept in the $e_i$=x' direction (TE mode), the scattered Raman signal at the output is analyzed varying the angle of the polarization $e_s$. For each output polarization, a spectrum is recorded, the background is subtracted and the Raman peaks are fitted with two Voigt functions. To correct the drift of the optical fibers during the experiment, a linear background is then subtracted from the Voigts areas in function of the scattered polarization $e_s$. The result presented figure \ref{fig:IRS} highlights clearly their periodic behavior. The two fitting components exhibit a 180$^\circ$ periodicity. The $\Delta \omega^{} = 0.84$ $\text{cm}^{-1}$ components is maximum at $e_s=90^\circ$ ($z^\prime$ direction) and the $\Delta \omega^{} = 3.08$ $\text{cm}^{-1}$ components at $e_s=0^\circ$ ($x^\prime$ direction), suggesting an attribution of the former frequency to the phonon 1 or 2 and the latter to the phonon 3 (see table \ref{tab:selection}. Assuming a degeneracy of the phonon 1 and 3, as it has been done in this paper, would change the shape of the polar representation of the component at $\Delta \omega^{} = 3.08$ $\text{cm}^{-1}$. This change is not observed here, mainly due to a limited resolution and too weak signal over noise ratio. Still, the results confirm the identification of two different phonons signatures in the Raman spectrum of IRS which demonstrate the potential of IRS.

We present in this work an experimental tool and method to partially resolved the effective stress tensor in integrated photonic structures, felt by the optical mode of the pump in telecom wavelength range. Similar results have been obtained between BRS and IRS configurations, demonstrating the robustness of the measurements. The three effective stress components considered in a triaxial model are then directly calculated, without fitting parameters, and present a good agreement with FEM simulations. Finally, the polarized-resolved version of the IRS is exploited to identify different phonon signatures in the system, revealing the potential of the technique. Further investigations, such as the study of the Raman signal as a function of the polarization state of the pump ($e_i$) would be interesting, as it would allow to characterize the stress in the waveguide for different electric field distribution (quasi-TE and quasi-TM optical mode), and therefore in different part of the waveguide.\\

\section*{Funding}
This project has received funding from the European Research Council (ERC) under the European Union’s Horizon 2020 research and innovation program (ERC POPSTAR - Grant Agreement No. 647342). We thank Mathieu Lancry for the interesting discussions regarding our results.

\section*{Disclosures}

The authors declare no conflicts of interest.

\bibliography{RAMAN}

\begin{thebibliography}{10}
\newcommand{\enquote}[1]{``#1''}

\bibitem{coudrain_towards_2012}
P.~Coudrain, J.~Colonna, C.~Aumont, G.~Garnier, P.~Chausse, R.~Segaud, K.~Vial,
  A.~Jouve, T.~Mourier, T.~Magis, P.~Besson, L.~Gabette, C.~Brunet-Manquat,
  N.~Allouti, C.~Laviron, S.~Cheramy, E.~Saugier, J.~Pruvost, A.~Farcy, and
  N.~Hotellier, \enquote{Towards efficient and reliable 300mm {3D} technology
  for wide {I}/{O} interconnects,} in \emph{2012 {IEEE} 14th {Electronics}
  {Packaging} {Technology} {Conference} ({EPTC}),}  (2012), pp. 330--335.

\bibitem{vianne_through-silicon_2015}
B.~Vianne, M.-I. Richard, S.~Escoubas, S.~Labat, T.~Schülli, G.~Chahine,
  V.~Fiori, and O.~Thomas, \enquote{Through-silicon via-induced strain
  distribution in silicon interposer,} {\protect\JournalTitle{Applied Physics
  Letters}} \textbf{106}, 141905 (2015).

\bibitem{fage-pedersen_linear_2006}
J.~Fage-Pedersen, L.~H. Frandsen, A.~V. Lavrinenko, and P.~I. Borel, \enquote{A
  {Linear} {Electrooptic} {Effect} in {Silicon}, {Induced} by {Use} of
  {Strain},} in \emph{3rd {IEEE} {International} {Conference} on {Group} {IV}
  {Photonics}, 2006.},  (2006), pp. 37--39.

\bibitem{de_wolf_relation_2015}
I.~De~Wolf, \enquote{Relation between {Raman} frequency and triaxial stress in
  {Si} for surface and cross-sectional experiments in microelectronics
  components,} {\protect\JournalTitle{Journal of Applied Physics}}
  \textbf{118}, 053101 (2015).

\bibitem{anastassakis_effect_1970}
E.~Anastassakis, A.~Pinczuk, E.~Burstein, F.~H. Pollak, and M.~Cardona,
  \enquote{Effect of static uniaxial stress on the {Raman} spectrum of
  silicon,} {\protect\JournalTitle{Solid State Communications}} \textbf{8},
  133--138 (1970).

\bibitem{wolf_stress_1996}
I.~D. Wolf, H.~E. Maes, and S.~K. Jones, \enquote{Stress measurements in
  silicon devices through {Raman} spectroscopy: {Bridging} the gap between
  theory and experiment,} {\protect\JournalTitle{Journal of Applied Physics}}
  \textbf{79}, 7148--7156 (1996).

\bibitem{ganesan_lattice_1970}
S.~Ganesan, A.~A. Maradudin, and J.~Oitmaa, \enquote{A lattice theory of
  morphic effects in crystals of the diamond structure,}
  {\protect\JournalTitle{Annals of Physics}} \textbf{56}, 556--594 (1970).

\bibitem{loechelt_measuring_1995}
G.~H. Loechelt, N.~G. Cave, and J.~Menéndez, \enquote{Measuring the tensor
  nature of stress in silicon using polarized off‐axis {Raman} spectroscopy,}
  {\protect\JournalTitle{Applied Physics Letters}} \textbf{66}, 3639--3641
  (1995).

\bibitem{poborchii_observation_2010}
V.~Poborchii, T.~Tada, and T.~Kanayama, \enquote{Observation of the forbidden
  doublet optical phonon in {Raman} spectra of strained {Si} for stress
  analysis,} {\protect\JournalTitle{Applied Physics Letters}} \textbf{97},
  041915 (2010).

\bibitem{poborchii_study_2007}
V.~Poborchii, T.~Tada, and T.~Kanayama, \enquote{Study of stress in a
  shallow-trench-isolated {Si} structure using polarized confocal near-{UV}
  {Raman} microscopy of its cross section,} {\protect\JournalTitle{Applied
  Physics Letters}} \textbf{91}, 241902 (2007).

\bibitem{claps_observation_2002}
R.~Claps, D.~Dimitropoulos, Y.~Han, and B.~Jalali, \enquote{Observation of
  {Raman} emission in silicon waveguides at 1.54 um,}
  {\protect\JournalTitle{Optics Express}} \textbf{10}, 1305--1313 (2002).

\bibitem{holmstrom_trace_2016}
S.~A. Holmstrom, T.~H. Stievater, D.~A. Kozak, M.~W. Pruessner, N.~Tyndall,
  W.~S. Rabinovich, R.~Andrew~McGill, and J.~B. Khurgin, \enquote{Trace gas
  {Raman} spectroscopy using functionalized waveguides,}
  {\protect\JournalTitle{Optica}} \textbf{3}, 891 (2016).

\bibitem{zhao_stimulated_2018}
H.~Zhao, S.~Clemmen, A.~Raza, and R.~Baets, \enquote{Stimulated {Raman}
  spectroscopy of analytes evanescently probed by a silicon nitride photonic
  integrated waveguide,} {\protect\JournalTitle{Optics Letters}} \textbf{43},
  1403--1406 (2018).

\bibitem{dhakal_nanophotonic_2016}
A.~Dhakal, P.~C. Wuytens, F.~Peyskens, K.~Jans, N.~L. Thomas, and R.~Baets,
  \enquote{Nanophotonic {Waveguide} {Enhanced} {Raman} {Spectroscopy} of
  {Biological} {Submonolayers},}  (2016).

\bibitem{wang_surface_2016}
Z.~Wang, M.~N. Zervas, P.~N. Bartlett, and J.~S. Wilkinson, \enquote{Surface
  and waveguide collection of {Raman} emission in waveguide-enhanced {Raman}
  spectroscopy,} {\protect\JournalTitle{Optics Letters}} \textbf{41}, 4146
  (2016).

\bibitem{wong_nanoscale_2018}
H.~M.~K. Wong, M.~K. Dezfouli, L.~Sun, S.~Hughes, and A.~S. Helmy,
  \enquote{Nanoscale plasmonic slot waveguides for enhanced {Raman}
  spectroscopy,} {\protect\JournalTitle{Physical Review B}} \textbf{98}, 085124
  (2018).

\bibitem{duverger_waveguide_1999}
C.~Duverger, J.~M. Nedelec, M.~Benatsou, M.~Bouazaoui, B.~Capoen, M.~Ferrari,
  and S.~Turrell, \enquote{Waveguide {Raman} spectroscopy: a non-destructive
  tool for the characterization of amorphous thin films,}
  {\protect\JournalTitle{Journal of Molecular Structure}} \textbf{480-481},
  169--178 (1999).

\bibitem{aggarwal_measurement_2011}
R.~Aggarwal, L.~Farrar, S.~Saikin, A.~Aspuru-Guzik, M.~Stopa, and D.~Polla,
  \enquote{Measurement of the absolute {Raman} cross section of the optical
  phonon in silicon,} {\protect\JournalTitle{Solid State Communications}}
  \textbf{151}, 553--556 (2011).

\bibitem{camassel_strain_2000}
J.~Camassel, L.~A. Falkovsky, and N.~Planes, \enquote{Strain effect in
  silicon-on-insulator materials: {Investigation} with optical phonons,}
  {\protect\JournalTitle{Physical Review B}} \textbf{63} (2000).

\bibitem{wolf_stress_1999}
I.~D. Wolf, \enquote{Stress measurements in {Si} microelectronics devices using
  {Raman} spectroscopy,} {\protect\JournalTitle{J. Raman Spectrosc.}} p.~8
  (1999).

\end{thebibliography}






\end{document}